\documentclass{article}

\usepackage{arxiv}

\usepackage[utf8]{inputenc} 
\usepackage[T1]{fontenc}    
\usepackage{hyperref}       
\usepackage{url}            
\usepackage{booktabs}       
\usepackage{amsfonts}       
\usepackage{nicefrac}       
\usepackage{microtype}      
\usepackage{lipsum}
\usepackage{graphicx}

\usepackage{balance}       
\usepackage{graphics}      
\usepackage[T1]{fontenc}   
\usepackage{txfonts}
\usepackage{wrapfig}
\usepackage{booktabs}
\usepackage{textcomp}
\usepackage{tabularx}
\usepackage{soul}
\usepackage{url}
\usepackage{cite}
\usepackage{multirow}
\usepackage{multicol}
\usepackage{xcolor}
\graphicspath{ {./images/} }

\title{User Concerns \& Tradeoffs in Technology-Facilitated Contact Tracing}

\author{
 Elissa M. Redmiles \\
 Max Planck Institute for Software Systems / Microsoft Research\\
  \texttt{eredmiles@gmail.com} 
}

\begin{document}
\maketitle
\begin{abstract}
    Please reference the published version of this paper which appears in Volume 2, Issue 1 of ACM Digital Government: Research and Practice (\url{https://dl.acm.org/doi/10.1145/3428093}.
\end{abstract}Technologists have proposed contact-tracing apps to help support public health during COVID19. In many Western nations there is the expectation that citizens will have the autonomy to decide whether or not to install these COVID19 apps. Yet, higher app adoption will have a greater the impact on public health~\cite{hinch2020effective, chan2020pact}, and thus technologists and public health officials are concerned about increasing adoption rates. 

How can we get users to install COVID19 apps? By understanding and mitigating the concerns and tradeoffs that serve as inputs to citizen's adoption decisions. This paper provides an empirically validated framework of the user-relevant components of COVID19 apps and corresponding user tradeoffs. Specifically, we enumerate user's considerations related to privacy -- including data collection, encryption, and data risks -- as well as the accuracy of these apps, their mobile costs, benefits, and transparency. We break down which tradeoffs are relevant for which contact tracing architectures (centralized vs. decentralized, location- vs. proximity-based). Additionally, we compare the relevant user considerations related to contact-tracing apps to alternative COVID19 technologies such as narrowcasts~\cite{chan2020pact}, location-personalized broadcasts of COVID19 hotspots. 

We validate our framework via both the literature and an empirical study of 1,0000 Americans, in which we decompose American's willingness (or unwillingness) to install COVID19 contact-tracing apps with different attributes.

While a heavy emphasis has been placed on privacy in the development and consideration of user reactions to contact-tracing apps~\cite{ahmed2020survey,sun2020vetting,Coronavi68:online,howgoodenough,li2020decentralized,simko2020covid,frimpong2020financial,Contactt3:online}, the framework and empirical evidence in this paper goes \textit{beyond}, but includes, privacy, to capture a broad spectrum of user considerations and potential adoption pitfalls for COVID19 technologies. 

\section{Technological Approaches to COVID19}
\label{sec:approaches}
There have been a number of technical approaches proposed to assist in the response to COVID19. Arguably the most discussed approach is digitally-enabled contact tracing. Contact tracing is a traditional epidemiological approach used to trace and limit the spread of infection~\cite{fairchild2007searching}. While technology-based contact-tracing approaches were proposed to address the 2014 Ebola outbreak, these approaches were never widely implemented~\cite{sacks2015introduction, danquah2019use}. Digital contact tracing via mobile app was first implemented at scale during the COVID19 pandemic.

These 'contact-tracing apps' are characterized by their data architecture: centralized or decentralized. In centralized contact-tracing apps, users are assigned an encrypted identifier by a trusted third party (TTP). Users' apps broadcast their identifier or a function of it to other apps within some distance \textit{d}. Users' apps then store lists of identifiers they have been in contact with, and if an app user reports that they have tested positive foor COVID19, their app will notify the TTP of it's stored list of contacts and the TTP will push a notification to the relevant, at-risk, app users. On the other hand, in decentralized contact-tracing apps, users' apps periodically generate \textit{anonymized identifiers} for them, which are broadcast to other apps within a given distance at periodic time intervals. Apps who's users have reported that they have tested positive for COVID19 push a list of exposed contact identifiers to a public list, rather than to a TTP as there is no TTP in a decentralized system. The other decentralized apps periodically pull this public list and check if they have any matches, if so, they notify the user that they have been exposed.

 The differences in the privacy guarantees offered by these two contact-tracing app architectures led to significant debates among experts regarding user privacy~\cite{ahmed2020survey,sun2020vetting,Coronavi68:online}. However, subsequent research on users' perceptions of the privacy these systems does not show a clear winner among end users, as we will discuss further in this paper~\cite{li2020decentralized}. 

contact-tracing apps are not the only technological responses proposed to COVID19. Another popular solution, sometimes combined with contact tracing functionality~\cite{15millio48:online}, are broadcast apps~\cite{raskar2020apps}. Some broadcast apps leverage existing public infrastructure that is normally used to provide emergency alerts to inform everyone in a given area of COVID19-related information. Others, such as ``narrowcasts''~\cite{chan2020pact}, provide personalized information, for example regarding COVID19 hotspot locations near the app user. Such narrowcasts can, similarly to contact-tracing apps, have centralized or decentralized architectures. In a centralized narrowcast architecture, the user's app pushes their location regularly to a  TTP. The TTP then pushes notifications of hotspots in the user's vicinity that it identifies based on the user's known location. In a decentralized narrowcast, a TTP publishes a list of hotspot locations. The user's app periodically pulls from this central list, matches the user's location (known only locally) against central list, and notifies the user of hotspots in user's vicinity.\footnote{For the least data collection, the user can also simply search their location on a map.}

Additional types of COVID19 response technologies include non-consensual surveillance of citizens using e.g., telecomm infrastructure~\cite{troncoso2020decentralized}, technologies to improve or scale manual contact tracing interviews~\cite{chan2020pact}, symptom checkers and self-diagnosis tools (e.g., thermometer checks)~\cite{covid19a46:online}, home health support for those diagnosed with COVID19~\cite{covid19a46:online}, and data-donation technologies for citizen scientists to contribute their data~\cite{CitizenS20:online}.

In this paper, we focus on user tradeoffs related to the most commonly proposed COVID19 technology: contact-tracing apps; as well as narrowcast apps, which have been frequently proposed or implemented as a feature of contact-tracing apps~\cite{15millio48:online}. 

\begin{table}[t]
\centering
\scriptsize
\begin{tabular}{rlll}
& \textbf{Component} & \textbf{Description} & \textbf{Dependencies}\\ 
  \hline
\parbox[t]{2mm}{\multirow{4}{*}{\rotatebox[origin=c]{90}{\textbf{Explicit}}}} & App Architecture & Type of app (e.g., centralized, decentralized, narrowcast) & -- \\
& Provider & Who is the app provider and/or the TTP (e.g., CDC, tech company) & --\\
& Data \& Accuracy& Data collected (e.g., proximity/location, health status) \& app accuracy & architecture \\
& Benefit: my health & My direct benefits (e.g., knowledge of my risk status, hotspots) & architecture, data collected \\
& Benefit: contacts' health & Direct benefits to contacts (e.g., they learn they are at risk) & architecture, data collected  \\
&&&\\
\hline
&&&\\
\parbox[t]{2mm}{\multirow{4}{*}{\rotatebox[origin=c]{90}{\textbf{Implicit}}}}& Information Protection & How the identifiers and/or data are protected (e.g., pseudoanonymized,  encrypted) & architecture\\
 & Privacy Costs & \textit{Who} can learn \textit{what} about me (e.g., neighbor can learn my health status) & architecture, data collected\\
& Mobile Costs & What will installing this app cost me? (e.g., battery life, GB of data used) & architecture (pull vs. push)\\ 
& User Agency & What do I control (data deletion, what I reveal) & architecture\\
&Benefit: environment safety & Others having app makes my environment safer & others actions\\
&Benefit: sense of altruism & Feeling like I helped others & \\
&Benefit: epidemiological data & e.g., decision makers know spread, infection rate & architecture\\
&&&\\
\hline
&&&\\
& Transparency & How transparent the app provider makes the implicit components&\\
&&&\\
\hline
\end{tabular}
\caption{Components of COVID19 apps.}
\label{tab:appcomponents}
\end{table}

\section{Potential Inputs to Users' COVID19 Adoption Decisions}
\label{sec:considerations}
In this section, we break down the attributes of COVID19 apps that are relevant to users' decisions regarding whether to adopt a COVID19 app (see Table~\ref{tab:appcomponents} for a summary of these attributes). We delineate users' potential considerations (e.g., concerns) related to these attributes, and provide empirical evidence supporting the relevance of these considerations on users' willingness to install. 

It is important to note that some app attributes that influence users' adopotion decisions are already transparent (e.g., the user can tell who is providing these app they download), while others (e.g., potential privacy risks) are not currently made transparent to the user by the app provider. In the absence of transparency, users may have their own, accurate or inaccurate, expectations about these implicit components, as we discuss below. 

Finally, it is important to note that this paper focuses on the features of COVID19 apps that users may consider in their adoption decision. This paper is not a comprehensive list of all possible user motivations to install COVID19 apps (or reasons to avoid them). Additional user-related factors that may affect willingness to install COVID19 apps (e.g., level of concern about COVID) are addressed briefly in Section~\ref{sec:controls}.

\textbf{Empirical Validation.} Where confirmed by external researchers, we cite the relevant research to support each user consideration detailed below. To validate considerations not already explored in prior work, we conducted our own empirical validation. To do so, we surveyed 1,000 Americans in June 2020. To maximize generalizability, we used quota sampling to ensure our respondents' demographics were representative of the U.S. in terms of age, race, education, income, gender, and geographic region. See the Appendix for more details regarding our survey methodology, including the limitations of our instrument. 

\subsection{App Provider}
There are many possible providers of COVID19 apps\footnote{List drawn in part from prior work~\cite{WillAmer45:online}.}: health protection agencies (e.g., CDC, FDA), who are required to be the providers of apps built on the Google-Apple exposure notification API~\cite{PrivacyP2:online}, health insurers~\cite{PayersAs70:online}, employers~\cite{Coronavi1:online}, technology companies~\cite{CitizenS70:online}, non-health agencies within a federal government or local government~\cite{Demonstr56:online}, non-profit organizations~\cite{Demonstr56:online}, international organizations such as the United Nations or World Health Organization~\cite{WillAmer45:online}, or universities~\cite{Contactt61:online}.

\textbf{Empirical Validation.} Prior work finds that users differ in their willingness to install COVID19 apps based on which entity provides the app, likely due to variance in their \textit{trust} of these entities or the contextual integrity of these entities collecting sensitive data related to health applications~\cite{WillAmer45:online, brown2020ethics,Whyproxi5:online}. 

\subsection{Accuracy} 
There are two types of errors that can occur in digital contact-tracing applications. The app could have false positives, in which the app notifies the user that they have been exposed to COVID19 when they actually have not been exposed. This can happen due to inaccuracies in proximity and location measurement or due to the app allowing non-validated self-reports of positive COVID19 status. The app could also have false negatives, where it fails to alert the user to all exposures to coronavirus.

\textbf{Empirical Validation.} Prior academic work has shown that accuracy affects users' reported willingness to install COVID19 apps~\cite{howgoodenough, frimpong2020financial}, and real world evidence from Google reviews and adoption statistics of released COVID19 apps further supports the relevance of accuracy considerations~\cite{NorthDak75:online}. 

\subsection{Data Collection}
There are two types of data that contact-tracing apps may collect: proximity data (who the app user has had contact with, where the "who" is anonymized as described in Section~\ref{sec:approaches}), or location data (where the app user has been)~\cite{ahmed2020survey}. 

\textbf{Empirical Validation.} Prior academic work shows that different users have different preferences for the type of data collected about them for contact tracing~\cite{li2020decentralized, simko2020covid}.

\subsection{User Agency}
Different COVID19 app implementations give users different levels of agency over their data. In the suggested implementations considered here, users always have the agency to decide whether to reveal their COVID-positive health status to an app, however, depending on app architecture, users may or may not have control over data retention. 

\textbf{Empirical Validation.} Prior work supports that users may have different preferences for the tradeoff between autonomy vs. decision-burden offered by apps with different implementations~\cite{li2020decentralized}. 

\subsection{Privacy Concerns}
\label{sec:privacy}
A contact-tracing app's architecture, including what data it collects and users' agency over that data, influences potential privacy concerns for the user. Privacy concerns can be thought of from a user lens as: ``\textit{who} can learn \textit{what} about me''.

There are five potential pieces of information (``whats'') that can be learned about a user: (1) That they are COVID-positive, (2) that they have been exposedd to COVID and are at risk of infection, (3) their social graph (i.e., who they have had contact with regardless of their health status), (4) their location information, (5) the COVID-status of their social group (e.g., other people of their race)\footnote{In a centralized system the TTP may be able to assign encrypted identifiers in such a way that it can track social groups~\cite{troncoso2020decentralized}}.

There are six potential attackers (``whos'') that can learn these pieces of information, depending on the app architecture. These attackers range from individuals to nation states, specifically: (1) other users of the app, (2) attackers who exploit the app (e.g., by placing Bluetooth beacons at specific locations, or who falsify the users' GPS coordinates), (3) the app (including individuals who work for the app), (4) any third-party service used by the app (including individuals who work for these services), (5) network providers, and (6) government entities that can use legal process to force the app to turn over data.

\begin{figure}
    \centering
    \includegraphics[width=0.8\textwidth]{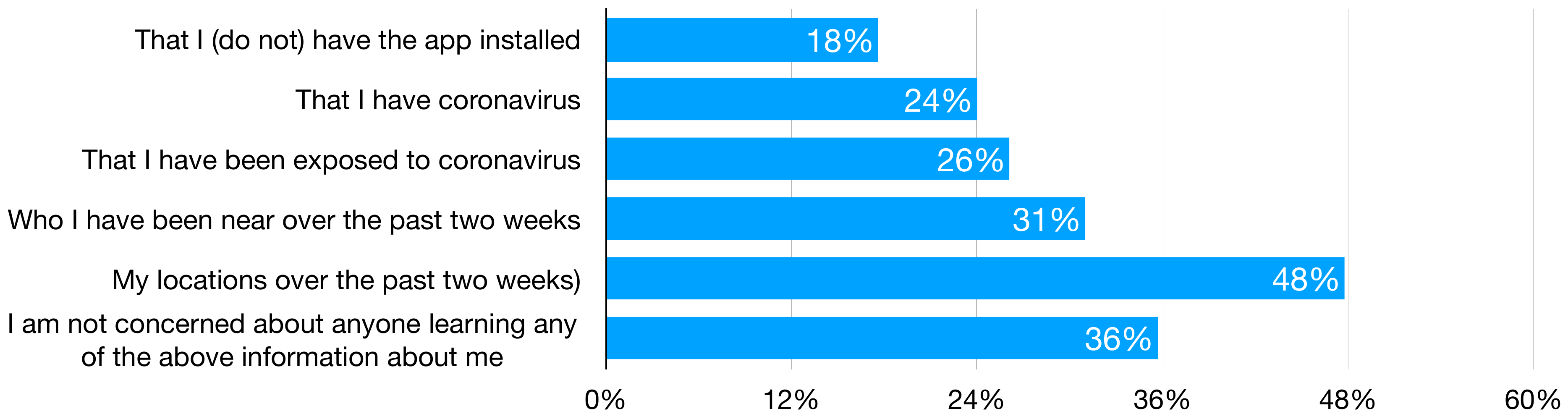}
    \caption{Survey responses from our survey of 1,000 Americans regarding their data privacy concerns regarding COVID19 apps (see Appendix for question wording).}
    \label{fig:sensitivedata}
\end{figure}

\begin{figure}
    \centering
    \includegraphics[width=0.8\textwidth]{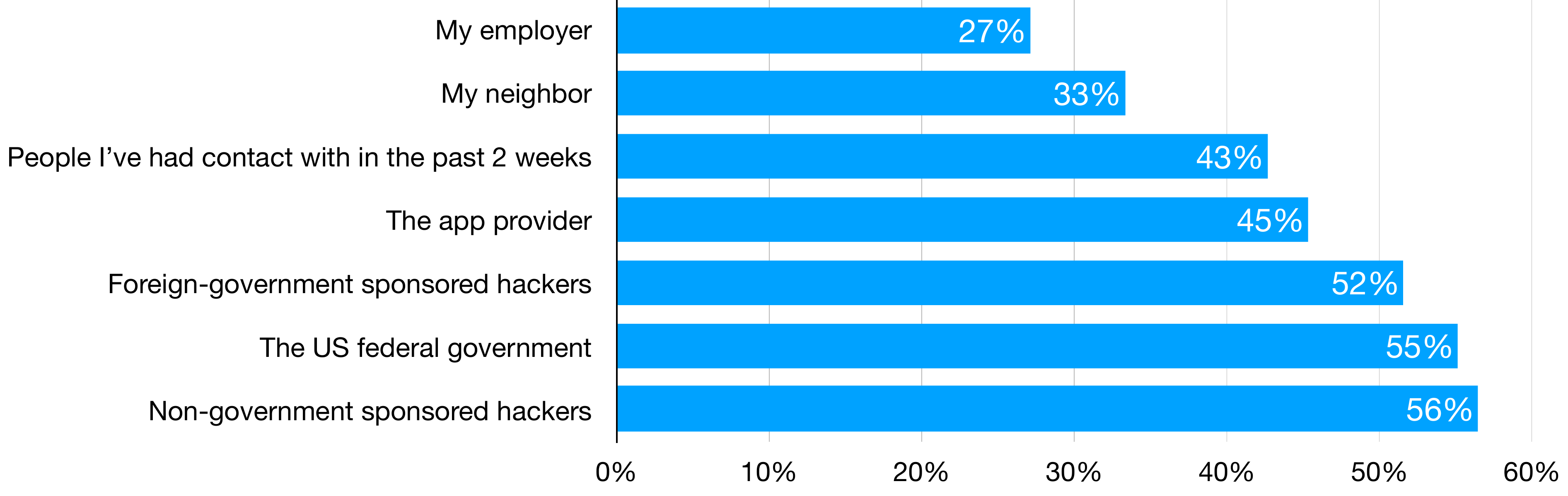}
    \caption{Entities about which potential COVID19 app users would be most concerned about learning each piece of information (Figure~\ref{fig:sensitivedata}) about which they were concerned. Responses are averaged across respondents with multiple data concerns.}
    \label{fig:sensitiveentities}
\end{figure}

\textbf{Empirical Validation.} Users have varying levels of concern about these pieces of information being leaked to different attackers. In open-answer questions, 20\% of our respondents volunteered that they were undecided or unwilling to install a COVID19-related app due to privacy or surveillance concerns. Additionally, we find differing levels of concern regarding the leakage of differing pieces of information (Figure~\ref{fig:sensitivedata} summarizes these perceptions), with the most (48\%) being concerned about someone being able to learn their location information and the least (18\%) being concerned about someone learning that they did / did not have the app installed. Respondents also varied in their concern regarding who could learn these pieces of information (Figure~\ref{fig:sensitiveentities}. The fewest respondents were concerned about their employer (27\%) or neighbor (33\%), who could e.g., place Bluetooth beacons at their place of work to eavesdrop on relevant information, being able to learn the information about which they were concerned while the most were concerned about hackers of various types being able to learn this information. Our results are confirmed by prior work by multiple other groups showing the relevance of privacy considerations in users' COVID19 app adoption decisions~\cite{howgoodenough,li2020decentralized,simko2020covid,frimpong2020financial,Contactt3:online}. 


\textbf{Potential for Inequity.} The potential for tracking COVID-status by social group may risk increased marginalization of underrepresented groups (as has already been a concern with high rates of COVID19 infection among communities of color~\cite{BlackAme57:online}), however documenting disparities can be critical to ensuring that marginalized groups receive the resources and support that they need~\cite{Howcanwe99:online}.

\subsection{Mobile Costs}
\label{sec:mobile}
contact-tracing apps are not cost-free, different app architectures may result in different mobile costs. For example, contact-tracing apps that rely on Bluetooth such as those built on the Google-Apple notification API~\cite{PrivacyP2:online} require the user to frequently use Bluetooth, which has known impacts on battery life~\cite{WhatYouS0:online}. Similarly, whether the app has push or pull architecture may also have implications for users' data costs (MB of mobile plan used for the app), storage costs (MB of space on the mobile phone used for the app), battery performance (impact on battery life from using the app), and the performance of other apps on their phone / their network speed.

\textbf{Empirical Validation.} Recent related work~\cite{trang2020one} and our own empirical work shows that mobile costs are an influential part of users' decisions to adopt COVID19 apps: twelve percent of our respondents reported that they were undecided or did not want to install a COVID19 app because they either do not use mobile apps as a habit or because of concerns about costs of mobile data use, limited phone storage, or battery life. In follow-up closed-answer questions asked to all respondents, regardless of their interest in installing a COVID19 app, we find that over half of respondents report that a noticeable impact on battery (66\%), storage (57\%), mobile data (62\%), or performance (64\%) costs would dissuade them from wanting to install a COVID19 app.

\textbf{Potential for Inequity.} Mobile costs are a potential source of inequity: less resourced users who are known to have less-featured / older mobile devices and are more likely to have limited mobile data~\cite{pew2019mobile,smith2015us,Washingt3:online}. These users may be disadvantaged by or unable to use apps with high mobile costs or whose functionality their devices do not support. 

\subsection{Benefits}
\label{sec:benefits}
COVID19 contact-tracing apps can have up to six different possible benefits depending on their architecture:
\begin{enumerate}
    \item Knowledge of risk: Users may feel that it is a benefit that the app can notify them if they have been exposed to someone who has COVID19.
       \item Knowledge of hotspots: Users may feel it is a benefit that they can use some contact-tracing apps to learn what locations near them have been visited by many people with COVID19~\cite{Newlocat33:online, chan2020pact}.
       \item Feeling of Altruism: Users may feel good about themselves for using or donating data through a contact-tracing  app~\cite{CitizenS20:online}, because they feel that they are helping society.
    \item Environment safety: Users may feel that by using a contact-tracing app they are improving the safety of their environment (e.g., country, community).
     \item Protecting loved ones: Users may feel that using the app helps them protect those they love or have come into contact with.
    \item Epidemiological data: Users may feel that a benefit of a contact-tracing app is that it allows for scientists or the government to collect COVID19-related data on infection rate (i.e., how many people are COVID19 positive) and/or spread (i.e., how many people have been exposed to COVID19).
    \end{enumerate}
 
The relevance of these benefits to users depends on: whether or not the user plans on taking action once they learn that they are at-risk (1,2); whether the user cares about the safety of those around them (3,4,5); whether the user thinks that others will take action once they learn that they are at risk (4); and whether the user believes that epidemiological data will be used / should be used to inform government/institutional COVID19 planning (e.g., lockdown lengths, PPE orders, hospital capacity planning) and whether the user cares about this planning (6).

\begin{figure}
    \centering
    \includegraphics[width=0.8\textwidth]{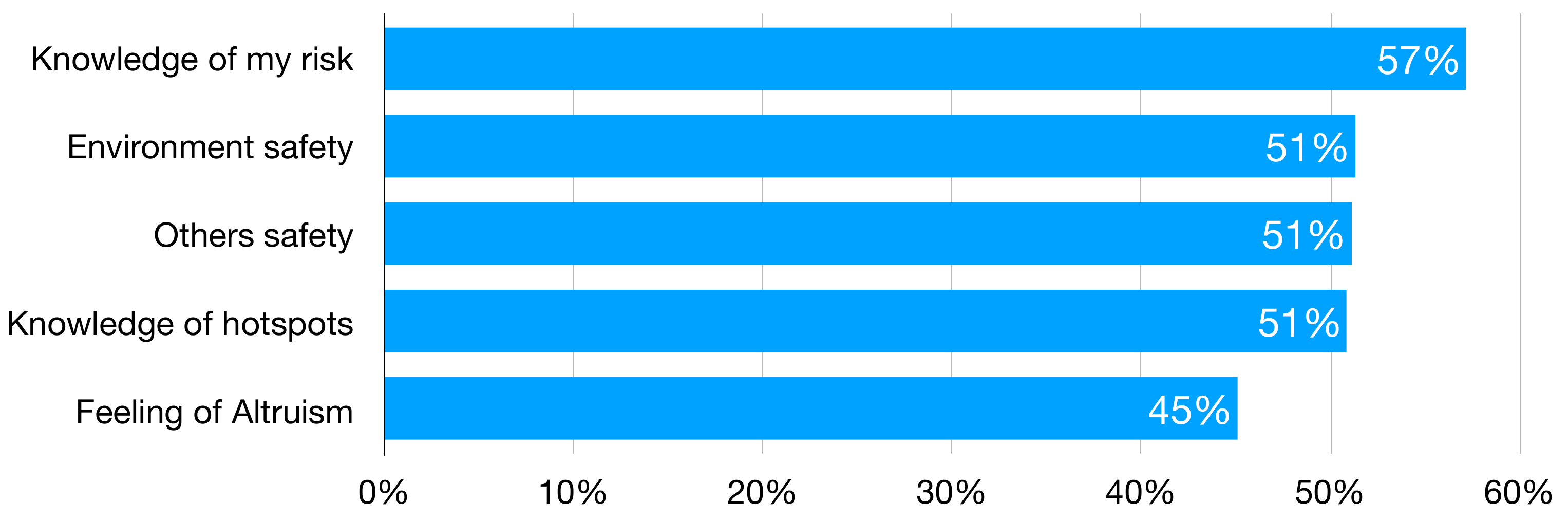}
    \caption{Proportion of survey respondents interested in installing a COVID19 app with different benefits. Responses are averaged across categories (see Appendix for exact survey wording).}
    \label{fig:benefits}
\end{figure}

\textbf{Empirical Validation.} Both prior work~\cite{trang2020one,li2020decentralized,CitizenS20:online} which covers a subset of these benefits, and our own work covering all six benefits shows that different benefits appeal to different users (Figure~\ref{fig:benefits}). Further, wide variation in reported willingness to adopt COVID19 apps in surveys that use differing descriptions of contact-tracing app benefits also supports the relevance of such benefits and how they are framed to users~\cite{Contactt3:online,WillAmer45:online,Washingt3:online}. 

It is important to note that COVID19 apps primarily benefit the health of others. Given that we tend to be self-focused~\cite{fehr2007human}, research is beginning to explore the use of direct (e.g., monetary) incentives for encouraging contact-tracing app adoption, as a supplement to the inherent societal benefits of these apps~\cite{frimpong2020financial}. 
    
\textbf{Potential for Inequity.} While desirable to users, hotspot features (benefit (2)) can negatively impact marginalized communities. Less resourced and minority communities have, thus far, experienced higher rates of COVID~\cite{BlackAme57:online}. Hotspot information may lead to increased marginalization of these communities and reduction in economic stimulus.

\section{COVID19 App Architecture Tradeoffs Through A User Lens}
\begin{figure}[h!]
    \centering
    \includegraphics[width=\textwidth]{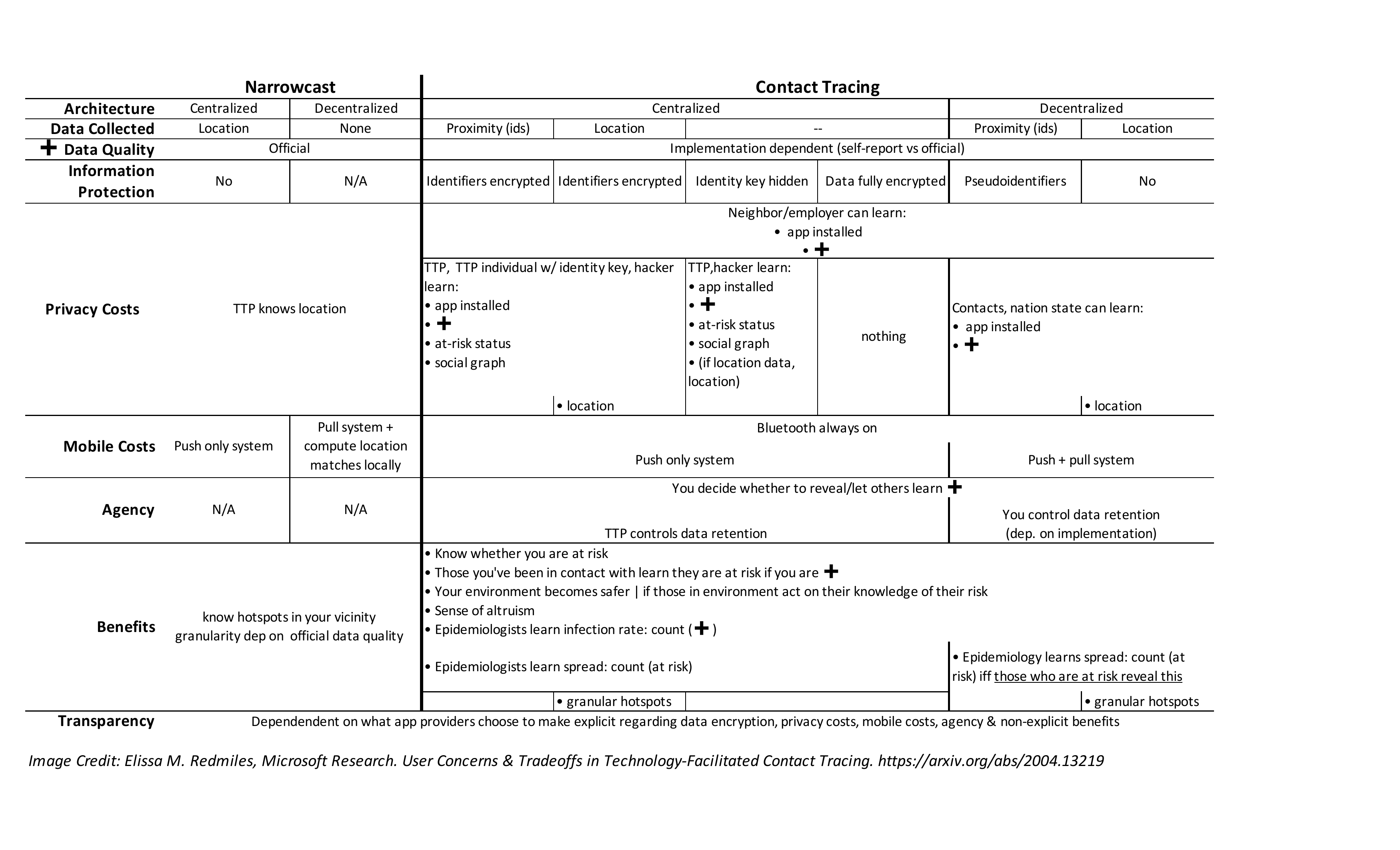}
    \caption{User-relevant tradeoffs between technologically-facilitated COVID19 approaches.}
    \label{fig:tradeofftable}
\end{figure}

Table~\ref{fig:tradeofftable} summarizes how the attributes that influence users' willingness to adopt COVID19 technologies detailed in the prior section vary based on functionality and architecture. In this section, we highlight the most critical differences in these architectures through a user lens.

\textbf{Broadcast vs. Contact Tracing.} Broadcast apps differ from contact-tracing apps in that they have lower privacy and mobile costs (they only know/can reveal a user's location and they do not require Bluetooth), however, they have fewer benefits both for users and for public health: they can only inform users of hotspots.

\textbf{Centralized vs. Decentralized Contact Tracing.} From a user perspective, these two competing contact tracing architectures differ in their privacy risks, mobile costs, and benefits. 

Centralized architectures allow the TTP, an individual at the TTP who knows the link between identifiers and real identity, and an attacker who hacks the TTP and the identifier link to learn whether you have the app installed, who is COVID-positive, who is at risk (COVID-exposed), and the social graph (who has had contact with whom) in the absence of testing COVID19 positive. Additionally, in centralized systems the user does not have agency over the deletion of their data. On the other hand, in  decentralized solutions the only thing that an attacker can learn is that the user's COVID-positive status and that they have the app installed (or not); the user has agency over their data in such systems as they can delete the app and data at any time. 

These systems may differ in mobile costs due to their differing push/push-pull architectures. Finally, they may also differ in the degree to which they benefit public health. In a centralized system the TTP can provide epidemiology data regarding spread (e.g., TTP can count the number of at risk persons), while in a decentralized system this information is available only if at risk people opt-in to sharing this data with an epidemiology server. 

\textit{Location vs. Proximity Data.} Finally, location-based contact-tracing apps of either architecture differ from proximity-based apps in that they have increased privacy risks (the user's location can be leaked) but also increased benefits: the app can provide hotspot information (benefit (2)), which many users find desirable (Figure~\ref{fig:benefits}). 

\section{Conclusion}
\label{sec:controls}
As shown in this paper, and through emerging evidence of adoption (or lack thereof) of contact-tracing apps throughout the world, there are a multitude of factors that must be considered when trying to encourage ethical adoption of digital contact-tracing apps. Beyond the app-specific factors covered in this work, user-specific factors such as sociodemographics, health status, type of employment (essential vs. non-essential worker), political leaning, and Internet skill may also influence people's willingness to adopt these technologies~\cite{li2020decentralized,Contactt3:online,WillAmer45:online,howgoodenough,simko2020covid}. Technologists should take care to consider both the app-specific considerations outlined in this paper and these additional user-specific considerations in their app architecture and marketing strategies, while researchers should continue to explore how contact-tracing app implementations may affect users' willingness to adopt and may exacerbate societal inequities.  

\section*{Acknowledgements}
With thanks to Paul England, Eszter Hargittai, Cormac Herley, Eric Horvitz, Gabriel Kaptchuk, Tadayoshi Kohno, Marina Micheli, Josh Rosenbaum, and Carmela Troncoso for feedback and conversations that contributed to this document.

\bibliographystyle{acm}
\bibliography{main}

\section*{Appendix}
\label{app:survey}
\subsection{Survey Methodology}
To validate the components of our empirical framework not already verified in prior work, we conducted a survey of 1,000 Americans in June 2020.

\textit{Survey Sampling.} To improve the generalizability of our results, we hired Cint, Inc. to conduct quota-based survey sampling to ensure that our survey respondents were representative of the demographics of the U.S. in terms of their age, race, education, income, gender, and geographic region.

\textit{Survey Questionnaire.}
Our survey questions were reviewed by at least two external survey experts, tested for respondent comprehension using cognitive interviews, and used multiple attention check questions to ensure respondent attentiveness, following survey methodological best practice~\cite{beatty2007research,redmiles2017summary}. 

We asked both open and closed-answer questions about respondents willingness to install COVID19 apps with particular functionalities or components. 
In the main body of the paper, we report the responses to these survey questions, in addition to referencing pre-existing empirical work, as  empirical validation of our framework.

To examine users' overall willingness to install COVID19 apps we asked the following questions:
    \begin{itemize}
        \item Would you be willing to install a coronavirus app? [Answer options: Yes, No, Unsure]
        \item Why [would you/would you not/are you not sure whether you would] be willing to install a coronavirus app? [Open answer]
    \end{itemize}
Respondents' answers to these questions were open-coded by two researchers, who met to ensure consistency on all responses. Results from these questions related to privacy are reported in Section~\ref{sec:privacy} and results related to mobile costs are reported in Section~\ref{sec:mobile}.
    
To understand what benefits (Section~\ref{sec:benefits}) would make users want to install a COVID19 app, we asked the following questions:
\begin{itemize}
    \item Now we would like to understand why you would want to install a coronavirus app. Please select all that apply. \\
    I would install a coronavirus app because I want to reduce...
    [answer options randomized except the final two]
    \begin{itemize}
        \item my risk of getting coronavirus
        \item my loved ones risk of getting coronavirus
        \item the risk of getting coronavirus for people I’ve been near
        \item the number of coronavirus cases in my area 
        \item Other: <text entry>
        \item I would never want to install a coronavirus app.
    \end{itemize}
	\item We’d like to understand a bit more about why you would want to install a coronavirus app. Please select all that apply.

	I would install a coronavirus app because I want to...
	[answer options randomized except the final two]
	\begin{itemize}
	    \item know which locations near me were recently visited by people who have coronavirus
\item help end the lockdown period sooner
\item meaningfully contribute to the effort to fight coronavirus
\item help researchers to get data on how many people have coronavirus
\item help researchers get data on how many people have been near someone who has coronavirus
   \item Other: <text entry>
        \item I would never want to install a coronavirus app.
	\end{itemize}
\end{itemize}

To examine the relevance of privacy costs (Section~\ref{sec:privacy}) to users' consideration of COVID19 apps, we asked the following questions: 
\begin{itemize}
    \item Using some coronavirus apps might allow someone to learn information about you. Which of the following types of information would you be worried about someone learning?  Please select all that apply. [answer choices randomized except the final choice]
    \begin{itemize}
\item That I have coronavirus
\item That I have been exposed to coronavirus
\item That I have the coronavirus app installed
\item My locations over the past two weeks (e.g., where your home is, what grocery store you visited)
\item Who I have been near over the past two weeks
\item I am not concerned about anyone being able to learn any of the above information about me.
\end{itemize}
\item \textit{Asked for each piece of information about which the respondent was concerned.} Please select which of these you are most concerned about being able to learn <that you have coronavirus/that you have been exposed...>. Please select all that apply.
\begin{itemize}
    \item My neighbor
    \item My employer
    \item People who have been near me in the past 2 weeks
    \item Whoever provides the app
    \item The US federal government 
    \item Non-government sponsored hackers
    \item Foreign-government sponsored hackers
    \item I am fine with all of the above learning that I have coronavirus
\end{itemize}

\end{itemize}

Finally, to examine the relevance of mobile costs to users' consideration of COVID19 apps we asked the following questions:
\begin{itemize}
\item Imagine that installing a coronavirus app will noticeably drain your phone’s battery. 
Would this stop you from installing the app? [Answer choices: Yes, No, Other <text entry>]
\item Imagine that installing a coronavirus app will use a noticeable amount of storage space on your mobile phone.
Would this stop you from installing the app? [Answer choices: Yes, No, Other <text entry>]
\item Imagine that installing a coronavirus app will use up a noticeable amount of your monthly mobile data.
Would this stop you from installing the app? [Answer choices: Yes, No, Other <text entry>]
\item Imagine that installing a coronavirus app will make using the other apps on your phone noticeably slower (e.g., it will take longer to perform an internet search).
Would this stop you from installing the app? [Answer choices: Yes, No, Other <text entry>]
\end{itemize}

\textit{Limitations.} As with all self-report studies, this empirical validation has inherent limitations due to potential response and generalizability biases. While we did our best to prevent such issues through careful respondent sampling and questionnaire design, our results should be interpreted with these limitations in mind.

\end{document}